\documentclass[prl, twocolumn,superscriptaddress,nobibnotes]{revtex4}
\usepackage{graphicx}
\usepackage{amssymb}
\usepackage{amsmath}
\usepackage{bm}
\usepackage{color}
\usepackage{float}
\usepackage{setspace}
\usepackage{physics}
\usepackage{subfigure}
\usepackage{soul}

\definecolor{orange}{RGB}{255,127,0}

\begin{document}

\title{Tilting flat bands in an empty microcavity}

\author{Ying Gao}\thanks{These authors contributed equally to this work.}

\affiliation{Institute of Molecular Plus, Tianjin University, Tianjin 300072, China} 
\author{Yao Li}\thanks{These authors contributed equally to this work.}

\affiliation{Institute of Molecular Plus, Tianjin University, Tianjin 300072, China} 
\author{Xuekai Ma}\thanks{Corresponding author: xuekaima@gmail.com}
\affiliation{Department of Physics and Center for Optoelectronics and Photonics Paderborn (CeOPP), Universit\"{a}t Paderborn, Warburger Strasse 100, 33098 Paderborn, Germany}

\author{Meini Gao}
\affiliation{Tianjin Key Laboratory of Low Dimensional Materials Physics and Preparing Technology, School of Science, Tianjin University, Tianjin 300072, China}

\author{Haitao Dai}
\affiliation{Tianjin Key Laboratory of Low Dimensional Materials Physics and Preparing Technology, School of Science, Tianjin University, Tianjin 300072, China}

\author{Stefan Schumacher}
\affiliation{Department of Physics and Center for Optoelectronics and Photonics Paderborn (CeOPP), Universit\"{a}t Paderborn, Warburger Strasse 100, 33098 Paderborn, Germany}
\affiliation{College of Optical Sciences, University of Arizona, Tucson, AZ 85721, USA}

\author{Tingge Gao}\thanks{Corresponding author: tinggegao@tju.edu.cn}
\affiliation{Institute of Molecular Plus, Tianjin University, Tianjin 300072, China} 

\begin{abstract}{Recently microcavities with anisotropic materials are shown to be able to create novel bands with non-zero local Berry curvature. The anisotropic refractive index of the cavity layer is believed to be critical in opening an energy gap at the tilted Dirac points. In this work, we show that an anticrossing between a cavity mode and a Bragg mode can also form within an empty microcavity without any birefringent materials. Flat bands are observed within the energy gap due to the particular refractive index distribution of the sample. The intrinsic TE-TM splitting and XY splitting induce the squeezing of the cavity modes in momentum space, so that the flat bands are spin-dependently tilted. Our results pave the way to investigate the spin orbit coupling of photons in a simple microcavity without anisotropic cavity layers.} 

\end{abstract}

\maketitle

\textit{Introduction.} The spin and orbit angular momentum always couple with each other in optically inhomogeneous medium, optical interface or anisotropic materials, resulting in the spin orbit coupling (SOC) of light \cite{SOI of light}. The SOC of light induces spin-dependent phenomena like the quantum spin-Hall effect of light \cite{Bliokh Science}, spin-Hall effect of light \cite{Geometrodynamics SOI}, spin-direction coupling of surface plasmons with photons \cite{SOI SPP}, transverse spin interaction between an optical fibre and a nanoparticle \cite{chiral nanowaveguide}, and the observation of a non-cyclic optical geometric phase during a non-Abelian evolution in a rolling-up microcavity \cite{SOI in asymmetric microcavity}. 

In optical microcavities, the cavity modes with TE or TM polarization are not degenerate at non-zero momenta, leading to the TE-TM splitting. The TE-TM splitting can also be described by a kind of photonic SOC \cite{Non Abelian TETM, Berry curvature in the microcavity}, which is closely related to the optical spin-Hall effect \cite{TETM, polariton spin hall effect, photon spin hall effect} and leads to the procession of the spin (pseudospin) of photon (polariton) modes in momentum space. Apart from the TE-TM splitting, the lower symmetry of the system produced during the growth of the sample creates another kind of energy splitting of the linearly polarized modes, i.e., XY splitting \cite{lagoudakis liquid crystal, review,  XY splitting}. 

With the TE-TM splitting, XY splitting, and an out-of-plane magnetic field, it is recently shown that two tilted Dirac points (also known as diabolic points) in momentum space of exciton polaritons can be opened and consequently show non-zero local Berry curvature \cite{Berry curvature polariton perovsktie, Berry curvature polariton GaAs}. The TE-TM splitting in a linearly polarized exciton polariton condensate in an anisotropic microcavity is shown to create a synthetic gauge field and excitation asymmetry against the cavity anisotropy axis \cite{polariton condensate synthetic, Non Abelian TETM}. In addition, an analogue of Rashba and Dresselhaus SOC due to the prominent TE-TM splitting and XY splitting can be realized in a microcavity filled with liquid crystal as the cavity layer where the refractive index can be tuned by an electric field \cite{liquid crystal_science, liquid crystal Light}. 

\begin{figure}[b]
 \centering
 \includegraphics[width=0.5\textwidth]{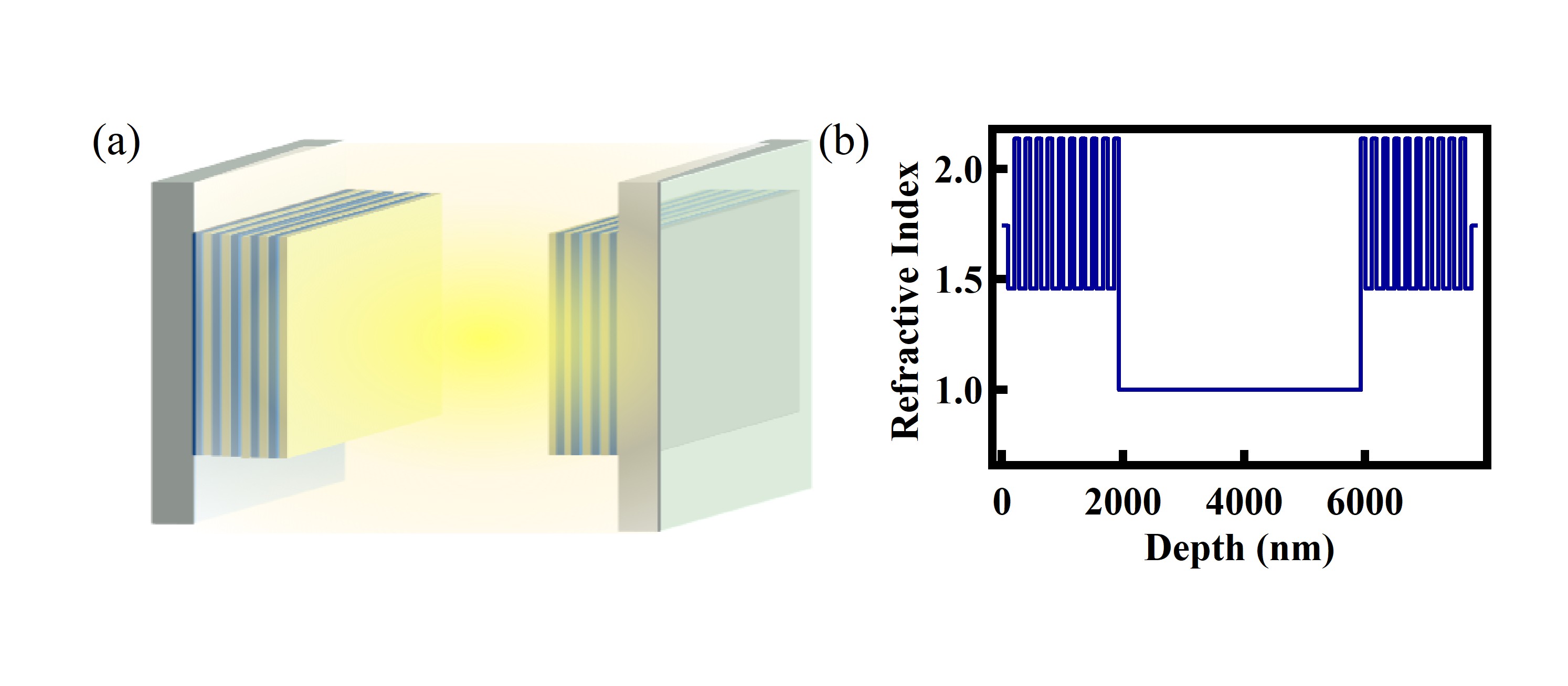} 
 \caption{\textbf{The microcavity.} (a) Schematics of the empty microcavity between two DBRs. (b) The refractive index distribution across the sample.} \label{fig:sample}
\end{figure}

The mode splittings in optical microcavities as mentioned above mainly focus on the strong light-matter coupling (e.g., exciton polaritons) or the coupling of different cavity modes with opposite parity. It is worth asking that whether the mode splitting can occur when the cavity mode and the Bragg mode in Distributed Bragg reflectors (DBRs) are brought into resonance. In this work, we realize such of a configuration in an empty optical microcavity and observe the mode splitting without anisotropic materials. Surprisingly, a flat band appears inside the opened energy gap and connects the anticrossing modes. Typically, flat bands are created in Lieb lattices or other periodic structures \cite{flat band-sutherland}. In this work the flat bands are observed in a microcavity which consists of an air layer and two DBRs. We demonstrate that the particular refractive index distribution of the system is the main reason for the creation of the flat bands. The flat bands in our system do not need complicated techniques to fabricate the photonic lattice structures. Thanks to the TE-TM splitting and XY splitting originating from the strain of the DBRs during the growth process, the spin polarized dispersion of the flat bands along specific direction in momentum space is not symmetric with respect to the normal incidence. The dispersion asymmetry of the flat bands can be swapped in the opposite spin polarization, which shares similarity with the topological edge modes in topological insulators. Our work offers a simple platform to investigate the SOC of photons without anisotropic layers in the cavity.

\textit{Principle of cavity-Bragg mode coupling.}
We firstly investigate the SOC of light in an empty optical microcavity formed between two DBRs. In such a configuration, the coupling of cavity modes and DBR Bragg modes, a small XY splitting due to the inhomogeneous strain generated during the fabrication of the sample, and the TE-TM splitting can simultaneously appear and influence the behavior of light. Accordingly, in the circular polarization basis of the wavefunction $|\Psi\rangle=[\Psi_\textup{C}^{+},\Psi_\textup{C}^{-},\Psi_\textup{B}^{+},\Psi_\textup{B}^{-}]^\textup{T}$, a $4\times4$ Hamiltonian can be used to describe this system, i.e.,
\begin{widetext}
\begin{equation}
H({\bf{k}})=
\begin{bmatrix}
E_\textup{C}+\dfrac{\hbar^2{\bf{k^2}}}{2m_\textup{C}} & \Omega_\textup{C}+\beta_\textup{C}{\bf{k^2}}e^{2i\varphi} & \gamma & 0 \\
\Omega_\textup{C}+\beta_\textup{C}{\bf{k^2}}e^{-2i\varphi} & E_\textup{C}+\dfrac{\hbar^2{\bf{k^2}}}{2m_\textup{C}} & 0 & \gamma \\
\gamma & 0 & E_\textup{B}+\dfrac{\hbar^2{\bf{k^2}}}{2m_\textup{B}} & \Omega_\textup{B}+\beta_\textup{B}{\bf{k^2}}e^{2i\varphi} \\
0 & \gamma & \Omega_\textup{B}+\beta_\textup{B}{\bf{k^2}}e^{-2i\varphi} & E_\textup{B}+\dfrac{\hbar^2{\bf{k^2}}}{2m_\textup{B}} \\
\end{bmatrix}.
\end{equation}
\end{widetext}
Here, $E_\textup{C}$ ($E_\textup{B}$) is the ground state of the cavity (Bragg) modes, $m_\textup{C}$ ($m_\textup{B}$) is the effective mass of the cavity (Bragg) modes, $\Omega_\textup{C}$ ($\Omega_\textup{B}$) denotes the strength of the XY splitting of the cavity (Bragg) modes due to the strain, $\beta_\textup{C}$ ($\beta_\textup{B}$) denotes the strength of the TE-TM splitting of the cavity (Bragg) modes, and $\varphi\in[0\ 2\pi]$ is the polar angle. $\gamma$ represents the coupling of the cavity modes and Bragg modes, giving rise to the anticrossing of them. The XY splitting together with the TE-TM splitting lead to the perpendicularly polarized TE and TM modes in the reciprocal space $(k_x, k_y)$ being squeezed along orthogonal directions, which results in their dispersions being crossed (at finite momentum) in one momentum direction, e.g. $k_x$ direction, and uncrossed in the perpendicular momentum direction, e.g. $k_y$ direction (see Fig. S1 in the SM).

\textit{Experimental realization.} In the experiments, we measure the dispersion of the empty microcavity, excited by white light from a halogen lamp in a hand-made momentum-space spectroscopy setup. The microcavity is formed by two DBRs (Fig.1(a)) which consist of alternative 10-period SiO$_2$/Ta$_2$O$_5$ layers. The cavity thickness is around 4 $\mu$m. We tilt one DBR so that the thickness of the cavity varies along the horizontal direction, in this way we can tune the cavity mode continuously. In the experiment we use a pinhole in the real space imaging plane which selects a small area in the pumping spot where the cavity length is constant. We directly measure the energy-momentum dispersion along $k_x$  direction and reconstruct the dispersion along $k_y$  direction by moving the last lens in front of the spectrometer. The polarization of the white light can be tuned to be linearly (circularly) polarized with a polarizing beam splitter and a half wave-plate (quarter wave-plate). The polarization of the reflected or transmitted light from the microcavity can be analyzed by a combination of a quarter wave-plate, half wave-plate, and linear polarizer. 

We focus on the cavity modes at the edge of the stop band, which come into resonance with the Bragg modes in our experiments. Due to the small refractive index of the air layer, the cavity modes have different dispersion curvatures (or effective masses) from the Bragg modes. At the intersecting points of the cavity modes and the Bragg modes, which act as optical analogue to the tilted Dirac points or diabolic points in the planar microcavities, we observe the anticrossing of the two modes, as demonstrated in Fig. 2(a) which shows the energy-momentum dispersion along $k_x$  direction when $k_y$ =0 under the excitation of unpolarized white light. The anticrossings are observed at the momentum of around 1.38 $\mu$m$^{-1}$ with the energy gap of around 0.022 eV between the cavity modes and the Bragg modes. The opened energy gaps are observed in both the TE and TM polarized modes with different momenta of around 1.36 $\mu$m$^{-1}$ and 1.43 $\mu$m$^{-1}$, respectively (see Fig. S2). Similar anticrossings of the cavity and Bragg modes can also occur at smaller energies (larger wavelengths) and larger momenta. An example can be found in Fig. S3. In our microcavity there is no any anisotropic material, thus the energy anticrossing should also be observed in other directions in the momentum space. The dispersion along $k_y$  direction (at $k_x$ =0) is demonstrated in Fig.  2(c) where the energy anticrossing appears at the momentum of around 1.38 $\mu$m$^{-1}$. The energy anticrossing between the cavity mode and Bragg mode in our microcavity excludes the necessity of anisotropic materials or circular birefringence, opening a new direction in the study of optical SOC based on the TE-TM splitting and XY splitting in a simple empty microcavity.  

\begin{figure}[t]
 \centering
 \includegraphics[width=0.5\textwidth]{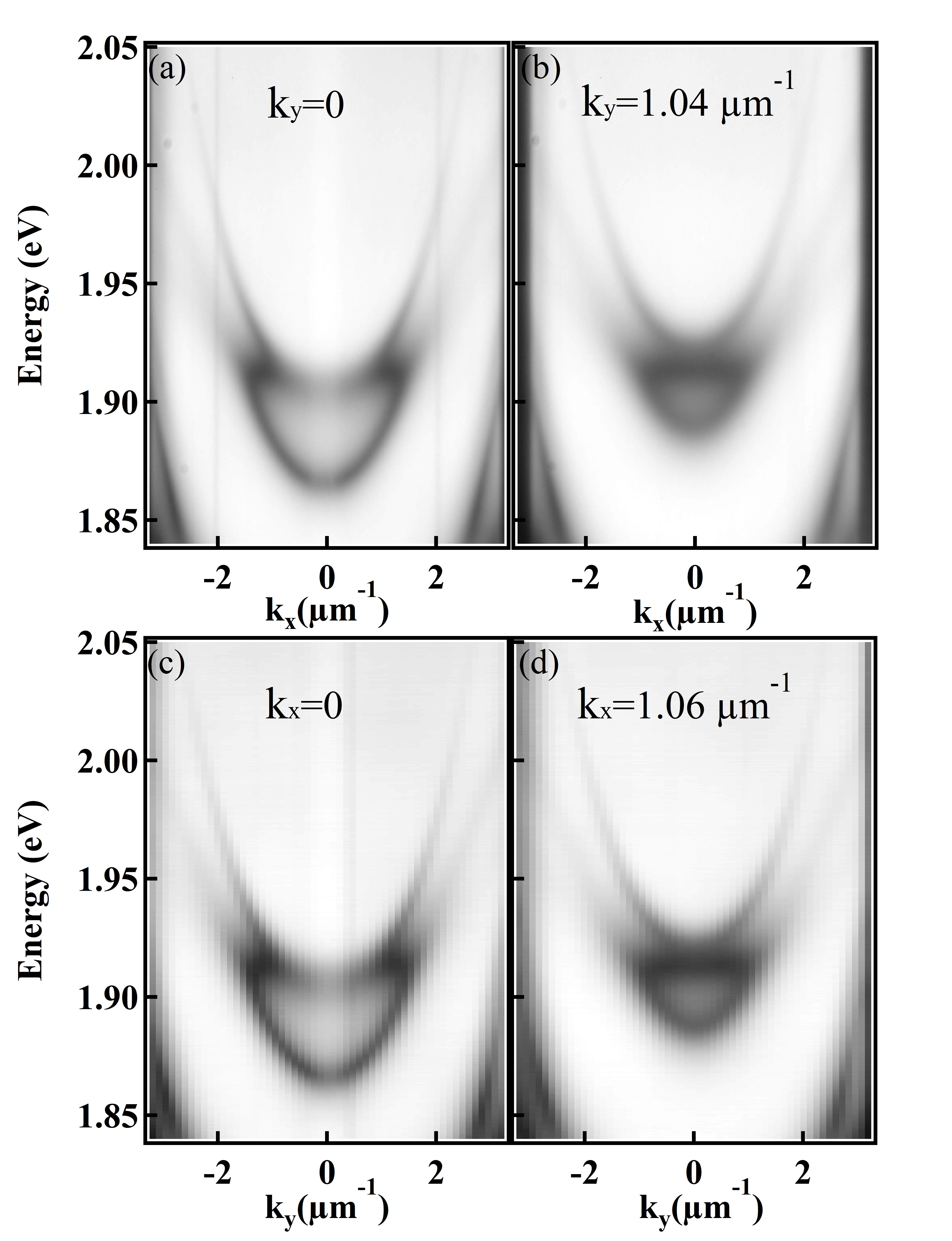} 
 \caption{\textbf{The dispersions of the microcavity.} The dispersions of the microcavity along $k_x$ direction at (a) $k_y$ =0 and (b) $k_y=$ 1.04 $\mu$m$^{-1}$. The dispersions of the microcavity along $k_y$ direction at (c) $k_x$ =0 and (d) $k_y$  direction at $k_x=$ 1.06 $\mu$m$^{-1}$. } \label{fig:dispersion}
\end{figure}

\textit{Flat bands.} Besides the energy anticrossing in the empty microcavity, we observe the flat bands connecting the two anticrossing bands, for example, when $k_y$  =0 as shwon in Fig. 2(a). The flat band can be seen more clearly when $k_y\neq0$, as shown in Fig. 2(b). The lineprofiles and fitted energy peaks measured at different momenta confirm the flatness of these bands (see Fig. S4 in the SM). The formation of the flat bands is due to the particular refractive index distribution across the sample, which can be confirmed by our numerical simulations based on the transfer matrix method (see Fig. S5 in the SM). The energy anticrossing and flat bands can be always observed as long as the cavity and Bragg modes are brought into resonance. Thanks to the wedge shape of the microcavity, we can tune the cavity mode continuously in the experiments. The energy anticrossing and flat bands can be observed periodically as the function of the cavity length (see Fig. S6 in the SM). We note that the refractive index of the cavity layer can be changed by adding the liquid crystal into the air layer, where the dispersion curvatures of the cavity mode and the Bragg mode become the same thus both the flat bands and the energy anticrossing disappear (see Fig. S7 in the SM).

\begin{figure}[t]
 \centering
 \includegraphics[width=0.5\textwidth]{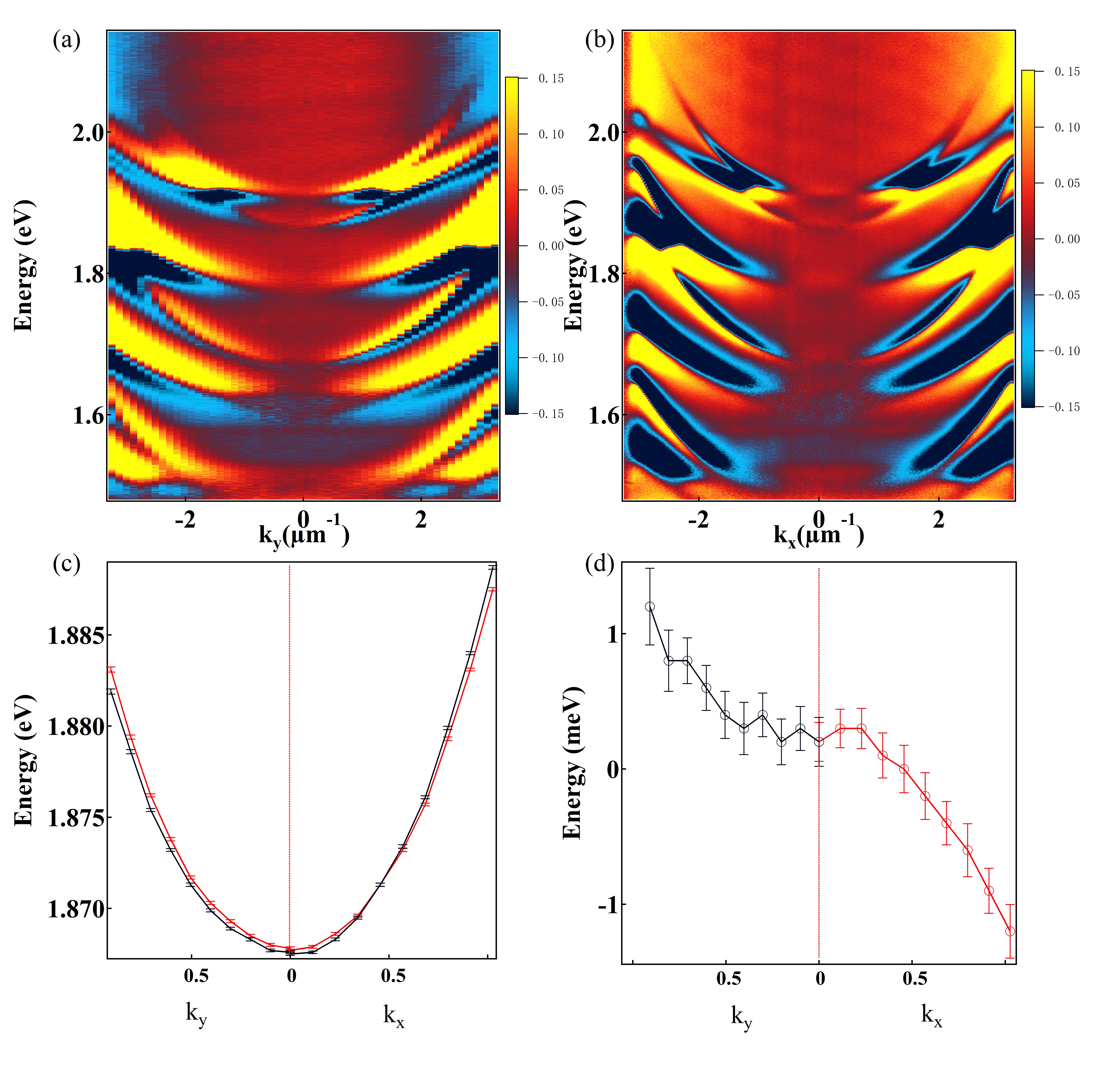} 
 \caption{\textbf{The band splitting of the microcavity.} Linearly polarized dispersions of the microcavity (a) along $k_y$  direction at $k_x$ =0 and (b) along $k_x$  direction at $k_y$  =0. (c) The fitted energy peak of the lower energy bands along $k_y$ ($k_x$ =0) and $k_x$ ($k_y$ =0)  directions. (d) The energy splitting along $k_y$ ($k_x$ =0) and $k_x$ ($k_y$  =0) directions.} \label{fig: spin polarized dispersion}
\end{figure}

\textit{Spin polarized bands.} The unintentional strain of the DBRs induces anisotropy into the system and reduces the cylindrical symmetry, leading to the Rashba-like non-Abelian gauge field \cite{non-abelian gauge fields}. In this case the bands will cross at specific positions in momentum space, as shown in Fig. 3(a, b). We fit the spectra of different linear polarizations along $k_x$ ($k_y$=0) and $k_y$ ($k_x$=0) directions and plot the energy peaks of the lower band of the anticrossing modes as shown in Fig. 3(c). The splitting between horizontal and vertical linear polarizations along $k_y$ and $k_x$ directions is shown in Fig. 3(d). These two bands cross each other along $k_x$ direction at 0.456 $\mu$m$^{-1}$ where the gauge field compensates for the effective magnetic field created by the TE-TM splitting. However, they never cross each other along $k_y$ direction at $k_x$=0. As a consequence, two linearly polarized modes in momentum space are squeezed along horizontal and vertical directions, respectively. Hence two circularly polarized modes are squeezed along diagonal and anti-diagonal directions, as shown in Fig. S8. 

\begin{figure}[t]
 \centering
 \includegraphics[width=\linewidth]{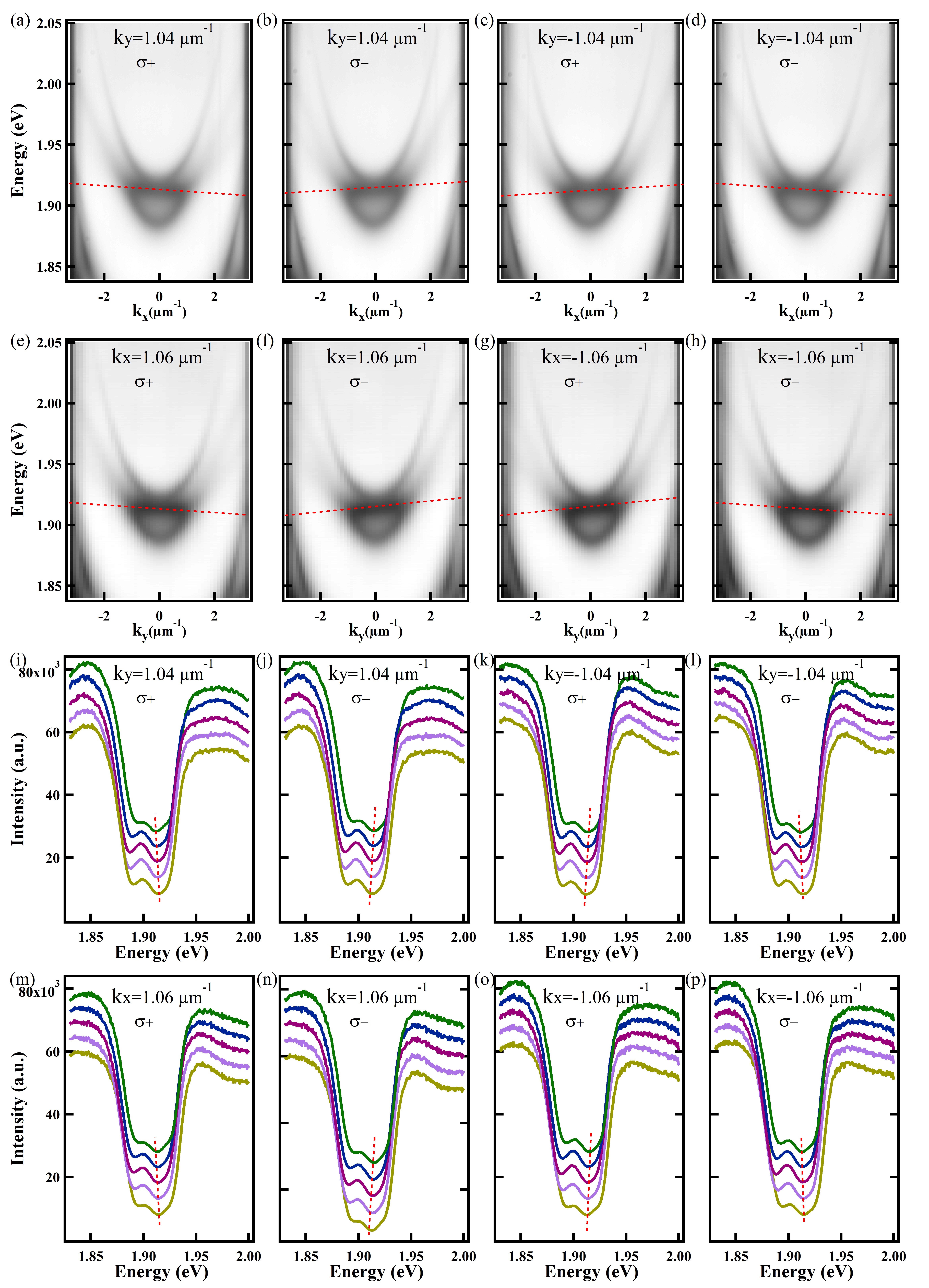} 
 \caption{\textbf{The spin polarized dispersions.} The right- ($\sigma^+$) and left-hand ($\sigma^-$) circularly polarized dispersions of the microcavity along $k_x$ direction at (a,b) $k_y$ = 1.04 $\mu$m$^{-1}$ and (c,d) $k_y$ is = -1.04 $\mu$m$^{-1}$. The right- ($\sigma^+$) and left-hand ($\sigma^-$) circularly polarized dispersions of the microcavity along $k_y$ direction at (e,f) $k_x$ = 1.06 $\mu$m$^{-1}$ and (g,h) $k_x$ = -1.06 $\mu$m$^{-1}$. (i-p) The lineprofiles taken at different momenta from (a-h), respectively.} \label{fig:spin polarized dispersion}
\end{figure}

In Fig. 4 we measure the dispersions of right-hand and left-hand circular polarization components along $k_x$ direction when $k_y\simeq$ 1.04 $\mu$m$^{-1}$. The two anticrossing bands and the flat bands are asymmetric with respect to the normal incidence. As shown in Fig. 4(a) the bands of the right-hand circular polarization are titled, such that the energy of the states at $k_x$\textgreater0 is smaller than the counterparts at $k_x$\textless0. Similar band asymmetry can be found in \cite{photonic gauge field} due to photonic gauge field in the structures. In our experiments, the bands of the left-hand circular polarization behave oppositely (Fig.  4(b)). This kind of anonymous results is swapped when the momentum $k_y$ is reversed, as shown in Fig. 4(c) and 4(d). The band-tilting can also be observed along $k_y$ direction when $k_x=$ $\pm$1.06 $\mu$m$^{-1}$, as shown in the Fig. 4 (e-h). The tilting of the flat bands (see the lineprofiles plotted in Fig. 4(i-p) and fitted energy peaks in Fig. S9) shows that different spin components gain opposite non-zero group velocities. The tilting of the flat bands at another position with different energies can be found in Fig. S10. 

Above results are measured under the excitation of unpolarized white light. If we modulate the polarization of the white light, the spin-dependent tilting of the flat bands can be reproduced. In Fig. S11 both the anticrossing bands and the flat bands are tilted in the same way as in Fig. 4(a, c, e, g) when the applied white light is right-hand circularly polarized, while the results are swapped and follow the trend in Fig. 4(b, d, f, h) when the white light is left-hand circularly polarized. It also excludes the possibility of any artifacts from the waveplates used in the optical detection setup. 

\textit{Summary.} To summarize, for the first time, we show that energy anticrossing can be observed in an isotropic microcavity without any anisotropic cavity medium, excluding the possibility of using any birefringent materials in the microcavity. The special refractive index distribution of the system gives rise to the flat bands. Thanks to the TE-TM splitting and XY splitting, the flat bands are tilted depending on the spin and the momentum of light. The tilting of the flat bands in different spin components shares the similarity with the edge modes in topological insulators and enables the manipulation of photonic modes in chip-based photonic devices at room temperature. Our results pave the way to investigate the SOC of light in a simple microcavity. 

\begin{acknowledgments}
TG acknowledges the support from the National Natural Science Foundation of China (NSFC, No. $11874278$). The Paderborn group acknowledges the Deutsche Forschungsgemeinschaft (DFG) through the collaborative research center TRR142 (project A04, grant No. 231447078).
\end{acknowledgments}



\begin{thebibliography}{99}

\bibitem{SOI of light} K. Y. Bliokh, F. J. Rodr{\'\i}guez-Fortu{\~n}o, F. Nori and A. V. Zayats, Nat. Photon. 9, 796 (2015).

\bibitem{Bliokh Science} Konstantin Y. Bliokh, Daria Smirnova and Franco Nori, Science 348, 1148 (2015).

\bibitem{Geometrodynamics SOI} Konstantin Y. Bliokh, Avi Niv, Vladimir Kleiner and Erez Hasman, Nature Photon. 2, 748 (2008).

\bibitem{SOI SPP} Jiao Lin, J. P. Balthasar Mueller, Qian Wang, Guanghui Yuan, Nicholas Antoniou, Xiao-Cong Yuan and Federico Capasso, Science 340, 331 (2013).

\bibitem{chiral nanowaveguide} Jan Petersen, J{\"u}rgen Volz and Arno Rauschenbeutel, Science 346, 67 (2014).

\bibitem{SOI in asymmetric microcavity} L.B. Ma, S.L. Li, V.M. Fomin, M. Hentschel, J.B. G{\"o}tte, Y. Yin, M.R. J{\"o}rgensen and O.G. Schmidt, Nat. Commun. 7, 1 (2016).

\bibitem{Berry curvature in the microcavity} O. Bleu, D. D. Solnyshkov, and G. Malpuech, Phys. Rev. B 97, 195422 (2018).

\bibitem{Non Abelian TETM} H. Ter{\c{c}}as, H. Flayac, D. D. Solnyshkov, and G. Malpuech, Phys. Rev. Lett. 112, 066402  (2014).

\bibitem{TETM} Alexey Kavokin, Guillaume Malpuech and Mikhail Glazov, Phys. Rev. Lett. 95, 136601 (2005).

\bibitem{polariton spin hall effect} C. Leyder, M. Romanelli, J. Ph. Karr, E. Giacobino, T. C. H. Liew, M. M. Glazov, A. V. Kavokin, G. Malpuech and A. Bramati, Nature Phys. 3, 628 (2007).

\bibitem{photon spin hall effect} Maria Maragkou, Caryl E. Richards, Tomas Ostatnick{\`y}, Alastair J. D. Grundy, Joanna Zajac, Maxime Hugues, Wolfgang Langbein, and Pavlos G. Lagoudakis, Opt. Lett. 36, 1095 (2011).

\bibitem{lagoudakis liquid crystal} P. Kokhanchik, H. Sigurdsson, B. Pi{\c{e}}tka, J. Szczytko, and P. G. Lagoudakis, Phys. Rev. B 103, L081406 (2021).

\bibitem{review} I A Shelykh, A V Kavokin, Yuri G Rubo, T C H Liew and G Malpuech, Semicond. Sci. Technol. 25 013001 (2010).

\bibitem{XY splitting} A. A. Demenev, Ya. V. Grishina, A. V. Larionov, N. A. Gippius, C.Schneider, S. H{\"o}fling, and V. D. Kulakovskii, Phys. Rev. B 96, 155308 (2017).

\bibitem{Berry curvature polariton perovsktie} L. Polimeno, M. De Giorgi, G. Lerario, L. De Marco, L. Dominici, V. Ardizzone, M. Pugliese, C. T. Prontera, V. Maiorano, A. Moliterni, C. Giannini, V. Olieric, G. Gigli, D. Ballarini, D. Solnyshkov, G. Malpuech, and D. Sanvitto, arXiv preprint arXiv: 2007. 14945 (2020).

\bibitem{Berry curvature polariton GaAs} A. Gianfrate, O. Bleu, L. Dominici, V. Ardizzone, M. De Giorgi, D. Ballarini, G. Lerario, K. West, L. Pfeiffer, D. Solnyshkov, D. Sanvitto, and G. Malpuech, Nature 578, 381 (2020).

\bibitem{polariton condensate synthetic} D. Biega{\'n}ska, M. Pieczarka, E. Estrecho, M. Steger, D. W. Snoke, K. West, L. N. Pfeiffer, M. Syperek, A. G. Truscott, and E. A. Ostrovskaya, arXiv preprint arXiv:2011.13290 (2020)

\bibitem{liquid crystal_science} Katarzyna Rechci{\'n}ska, Mateusz Kr{\'o}l, Rafa{\l} Mazur, Przemys{\l}aw Morawiak, Rafa{\l} Mirek, Karolina {\L}empicka, Witold Bardyszewski, Micha{\l} Matuszewski, Przemys{\l}aw Kula, Wiktor Piecek, Pavlos G. Lagoudakis, Barbara Pi{\c{e}}tka, Jacek Szczytko, Science 366, 727 (2019).

\bibitem{liquid crystal Light} Katarzyna Lekenta, Mateusz Kr\'{o}l, Rafał Mirek, Karolina Łempicka, Daniel Stephan, Rafał Mazur, Przemysław Morawiak, Przemysław Kula, Wiktor Piecek, Pavlos G. Lagoudakis, Barbara Pi{\c{e}}tka and Jacek Szczytko, Light: Science \& Applications 7, 1 (2018).

\bibitem{flat band-sutherland} Bill Sutherland, Phys.Rev.B 34, 5208 (1986).

\bibitem{photonic gauge field} Yaakov Lumer, Miguel A. Bandres, Matthias Heinrich, Lukas J. Maczewsky, HananHerzig - Sheinfux, Alexander Szameit and Mordechai Segev, Nat. Photon. 13, 339 (2019).

\end{thebibliography}
\end{document}